\documentclass[conference, a4paper]{IEEEtran}
\IEEEoverridecommandlockouts

\usepackage[a4paper, left=1.42cm, right=1.42cm, top=1.9cm, bottom=4.25cm, columnsep=0.25in]{geometry}

\usepackage{newtxtext,newtxmath}
\usepackage[T1]{fontenc}

\usepackage{algorithm} 
\usepackage{algpseudocode} 

\usepackage{cite}
\usepackage{amsmath} 
\usepackage{graphicx}
\usepackage{textcomp}
\usepackage{xcolor}
\usepackage{array}
\usepackage{booktabs}
\usepackage{multirow}
\usepackage{tabularx}
\usepackage{url}
\usepackage{stfloats}
\usepackage{subcaption}
\usepackage{enumitem}
\usepackage{verbatim}
\usepackage{caption}
\usepackage{cleveref}

\def\BibTeX{{\rm B\kern-.05em{\sc i\kern-.025em b}\kern-.08em
    T\kern-.1667em\lower.7ex\hbox{E}\kern-.125emX}}
\begin{document}

\title{CovertComBench: A First Domain-Specific Testbed for LLMs in Wireless Covert Communication\\
}
\author{Zhaozhi~Liu,~Jiaxin~Chen,~Yuanai~Xie*,~Yuna~Jiang,~Minrui~Xu, Xiao~Zhang, Pan~Lai, Zan~Zhou

\thanks{Z.~Liu,~J.~Chen,~Y.~Xie, X.~Zhang, P.~Lai are with the School of Computer Science, South-Central Minzu University, Wuhan 430074, China. Emails: 2024120415@mail.scuec.edu.cn, 2024120388@mail.scuec.edu.cn, 2023002@scuec.edu.cn,  xiao.zhang@my.cityu.edu.hk, plai1@ntu.edu.sg.}
\thanks{Y.~Jiang is with the School of Communications and Information Engineering, Nanjing University of Posts and Telecommunications, Nanjing 210042, China. Email:yunajiang@njupt.edu.cn.}
\thanks{M.~Xu is with the College of Computing and Data Science, Nanyang Technological University, Singapore. Email: minrui001@e.ntu.edu.sg.}
\thanks{Z.~Zhou is with the State Key Laboratory of Networking and Switching Technology, Beijing University of Posts and Telecommunications, Beijing 100876, China (email: zan.zhou@bupt.edu.cn).}
\thanks{J.~Kang is with the School of Automation, Guangdong University of Technology, Guangzhou 510006, China. Email: kavinkang@gdut.edu.cn.}
}
\maketitle

\begin{abstract}
The integration of Large Language Models (LLMs) into wireless networks presents significant potential for automating system design. However, unlike conventional throughput maximization, Covert Communication (CC) requires optimizing transmission utility under strict detection-theoretic constraints, such as Kullback-Leibler divergence limits. Existing benchmarks primarily focus on general reasoning or standard communication tasks and do not adequately evaluate the ability of LLMs to satisfy these rigorous security constraints. To address this limitation, we introduce CovertComBench, a unified benchmark designed to assess LLM capabilities across the CC pipeline, encompassing conceptual understanding (MCQs), optimization derivation (ODQs), and code generation (CGQs). Furthermore, we analyze the reliability of automated scoring within a detection-theoretic ``LLM-as-Judge'' framework. Extensive evaluations across state-of-the-art models reveal a significant performance discrepancy. While LLMs achieve high accuracy in conceptual identification (81\%) and code implementation (83\%), their performance in the higher-order mathematical derivations necessary for security guarantees ranges between 18\% and 55\%. This limitation indicates that current LLMs serve better as implementation assistants rather than autonomous solvers for security-constrained optimization. These findings suggest that future research should focus on external tool augmentation to build trustworthy wireless AI systems.
\end{abstract}


\section{Introduction}
Driven by sustained breakthroughs, Large Language Models (LLMs)~\cite{brown2020language, xu2024cached, xu2024large} have rapidly extended their capabilities beyond traditional Natural Language Processing (NLP) tasks and are now being applied across a wide range of scientific and engineering domains. In wireless communications, LLMs have demonstrated remarkable potential in automating system modeling, optimizing resource allocation, and assisting with code-level implementation~\cite{lee2024allocation, yang2025wirelessgpt, li2025graph}. Recent works~\cite{mekrache2024intent, tong2024connectgpt, kan2024mobile} have showcased the application of LLMs in wireless communications, including intent-based management of next-generation networks, integration with connected and automated vehicles, and instruction fine-tuning for 5G network analysis, marking a significant step toward intelligent network and service management in wireless communication systems.
Among emerging directions in wireless communications, covert communication (CC) has garnered increasing attention due to its crucial role in ensuring information security and privacy. Unlike conventional communication paradigms that prioritize throughput or reliability, CC aims to conceal the very existence of communication under an adversary's surveillance. This unique objective imposes a strict covertness constraint, leading to a fundamental trade-off between maximizing the legitimate user's transmission rate and minimizing detectability by the warden~\cite{chen2023covert}. Consequently, optimization in CC involves intricate reasoning grounded in statistical detection theory and probabilistic modeling, which poses substantial challenges even for advanced LLMs.

In light of these challenges, it is essential to assess whether LLMs can effectively reason about the mathematical formulations, optimization objectives, and engineering principles that define CC systems. However, existing LLM benchmarks~\cite{11021474teleqa, li2025wirelessmathbench, 10975994oran } primarily focus on general reasoning or conventional communication tasks, where covertness is not explicitly modeled as a hard constraint. To date, no specialized benchmark exists for evaluating LLM performance in CC, leaving a significant gap in assessing their capability to handle security-sensitive wireless problems.

To bridge this gap, we propose \textbf{CovertComBench}\footnote{https://huggingface.co/datasets/CovertComBench/CovertComBench}, the first systematic benchmark designed to evaluate the reasoning and problem-solving capabilities of LLMs in CC. CovertComBench is a 100\% human-verified benchmark characterized by vertical domain depth and strict adherence to the physical constraints of CC, prioritizing reasoning density over dataset magnitude. The benchmark covers a wide range of modern communication system models~\cite{qu2025mobile}, including Intelligent Reflecting Surface (IRS), Non-Orthogonal Multiple Access (NOMA), and Multiple-Input Multiple-Output (MIMO) systems, and spans three distinct task categories: Multiple-Choice Questions (MCQs), Optimization Derivation Questions (ODQs), and Code Generation Questions (CGQs). Furthermore, we design a multi-dimensional evaluation framework that incorporates both human expert assessment and an LLM-as-Judge (LAJ) mechanism~\cite{zheng2023judging} to ensure reliable and transparent evaluation. By statistically analyzing the discrepancy between human and LLM-based scores, we quantify the reliability of automated evaluation in domain-specific benchmarks.

Our extensive experiments across multiple state-of-the-art models reveal that while LLMs perform strongly on conceptual and coding tasks, they exhibit a sharp decline in performance on multi-step mathematical reasoning tasks. This finding exposes a key limitation in current LLMs’ ability to autonomously solve higher-order optimization problems. It also highlights the potential of integrating LLMs with external symbolic computation tools (e.g., SymPy, Mathematica) to enhance their reasoning capabilities.

In summary, our main contributions are as follows: \begin{itemize} \item \textbf{First Comprehensive Benchmark for CC:}We introduce \textit{CovertComBench}, the first benchmark dedicated to evaluating LLMs in CC, encompassing a broad spectrum of rigorously validated tasks across multiple system models.
\item \textbf{Novel Multi-dimensional Evaluation Framework:} We design a structured evaluation system that assesses LLMs across tasks spanning conceptual understanding, mathematical derivation, and code implementation. 
\item \textbf{Benchmarking the Evaluator:} We quantify the reliability of the LLM-as-a-Judge mechanism in CC. By comparing model scores with expert evaluations, we identify reliability issues in the tested models. 
\item \textbf{In-depth Empirical Findings:} We provide comprehensive insights into the strengths and limitations of LLMs in CC reasoning, offering empirical evidence to guide future research in LLM-assisted wireless optimization. \end{itemize}

\section{The CovertComBench}
This section details the CovertComBench, covering its design motivation, construction pipeline, and dataset structure, and concludes with a statistical analysis.

\begin{figure*}[ht]
    \centering
    \includegraphics[width=0.82\linewidth]{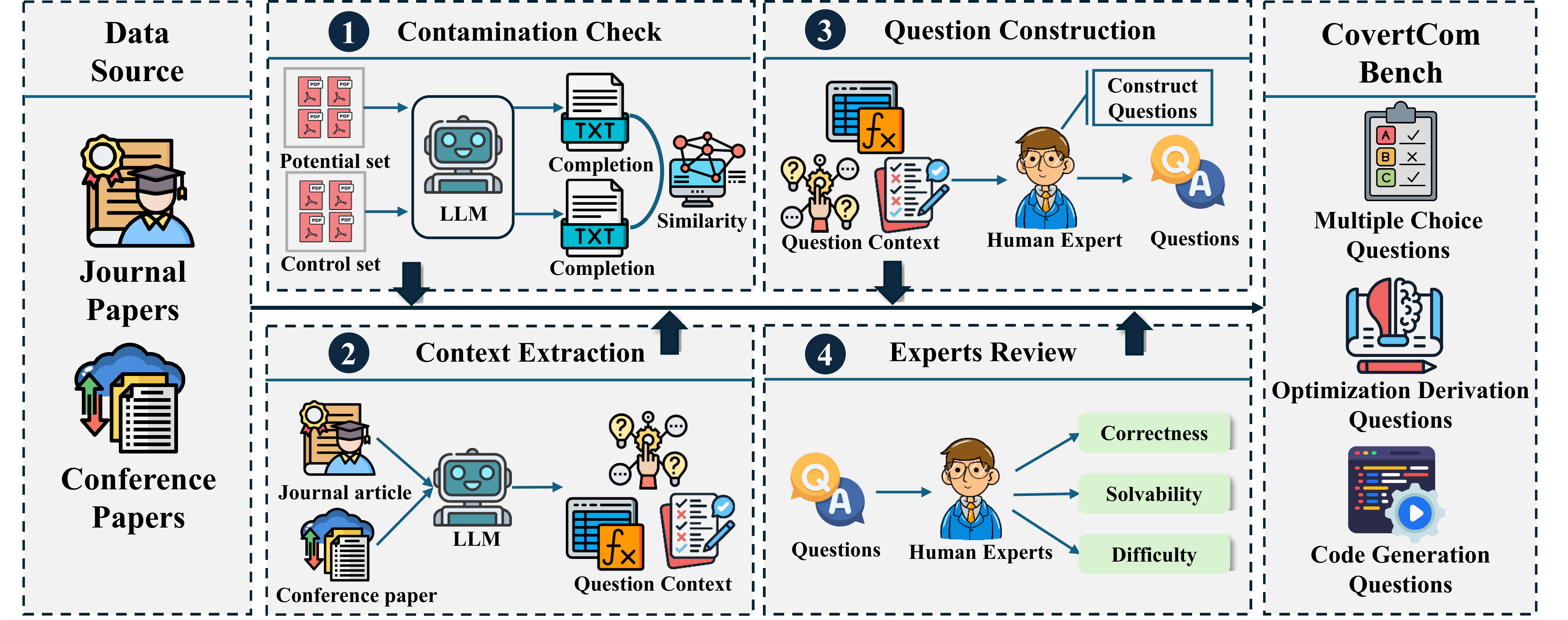}
    \caption{The construction pipeline of CovertComBench. The four main stages are: (1) Contamination Check, (2) Context Extraction, (3) Question Construction, and (4) Experts Review.}
    \label{fig:covert_com_bench_pipeline}
\end{figure*}

\subsection{Motivation}
The essence of CC lies in the fundamental trade-off between covertness and covert rate, a challenging optimization problem constrained by noise power and detection strategies. Maximizing the covert rate while remaining undetectable is inherently a challenging mathematical optimization problem. The emergence of LLMs offers a novel perspective: serving as “domain problem-solvers” that translate natural language descriptions of CC scenarios into formal optimization problems, automatically formulating objective functions and constraints, and even deriving solutions directly. However, the engineering capabilities of LLMs in such highly specialized tasks remain largely unexplored. How do different LLMs perform in terms of modeling accuracy, reasoning depth, and optimization effectiveness? At present, the absence of a standardized evaluation framework leaves these questions unanswered, making the construction of a domain-specific benchmark both necessary and urgent.

\subsection*{Problem Formulation}
Our work aims to evaluate the problem-solving capabilities of large language models within the context of CC. We formalize this capability as a constrained optimization problem centered on maximizing task-specific utility under a strict covertness constraint. A canonical CC scenario involves a sender ($S$), a legitimate receiver ($R$), and a warden ($W$) who monitors the channel to detect any unauthorized transmission. The goal of $S$ is to deliver information or accomplish a task for $R$ while remaining undetected by $W$. We denote by $Y_R$ and $Y_W$ the random variables observed by the receiver and the warden, respectively. 

For any given instance $x$ from CovertComBench, the model is required to generate an answer $a$ from the candidate space $\mathcal{A}$ to maximize a task-related utility function $U(a|x)$, subject to a covertness constraint that limits the probability of detection by the warden to a predefined threshold $\alpha \in [0, 1]$. The optimization problem can thus be formally stated as:
\begin{equation} \label{eq:optimization_problem}
\max_{a \in \mathcal{A}} U(a|x) 
\quad \text{s.t.} \quad 
\Pr\{\text{detected}(a, x)\} \leq \alpha
\end{equation}

To instantiate the binary event $\text{detected}(a, x)$ in Eq.~\eqref{eq:optimization_problem}, we adopt the \textit{Kullback--Leibler (KL) divergence}, a standard measure of statistical distinguishability in CC. Specifically, $\text{detected}(a, x)$ is defined as $\text{KL}(P_w \parallel P_a) > \varepsilon$, where $P_w$ denotes the warden’s observed signal distribution when no CC occurs (the null hypothesis), $P_a$ denotes the observed distribution when the model generates answer $a$ for instance $x$ (the alternative hypothesis), and $\varepsilon$ is a predefined detection threshold. A smaller $\varepsilon$ enforces stricter covertness by constraining the divergence between $P_w$ and $P_a$. This probability is determined by the hypothesis testing model and parameters defined in each instance’s metadata.

The three question types in CovertComBench, i.e., MCQs, ODQs, and CGQs, represent distinct instantiations of this general utility function $U(\cdot)$, designed to probe different facets of a model's capabilities:

\begin{itemize}
    \item \textbf{MCQs :} These correspond to identification tasks. The utility function is a specialized binary case where $U(a|x) = 1$ if the model's answer $a$ is the correct option, and $0$ otherwise.

    \item \textbf{ODQs:} These assess the model's reasoning and derivation abilities. The utility function $U(\cdot)$ is implemented as a procedural, rubric-based score, awarding a value between 0 and 1 based on the correctness, completeness, and clarity of the derivation steps.

    \item \textbf{CGQs:} These measure the model's programmatic construction skills. The utility function $U(\cdot)$ is a composite metric that jointly evaluates functional correctness, resource efficiency, and adherence to specific covert design patterns in the generated code or configuration.
\end{itemize}

Through this formulation, CovertComBench provides not merely a collection of problems, but a standardized testbed for evaluating a model's core utility under formal, quantifiable covertness constraints.

\subsection{CovertComBench Construction Pipeline}

To systematically evaluate a model's ability to solve the constrained optimization problem defined in the previous section, we constructed CovertComBench. As illustrated in \Cref{fig:covert_com_bench_pipeline}, its construction follows a rigorous four-stage pipeline designed to generate a high-quality and reproducible benchmark embodying this formal problem. This pipeline consists of S1: Source Vetting and Decontamination, S2: Context Extraction and Question Formulation, S3: Instance Refinement and Expert Validation, and S4: Finalization, Trial Run, and Baselining. 
Each stage ends with a dedicated Quality Gate (QG1–QG4) to ensure output quality, as detailed in \Cref{alg:covertcombench_pipeline}.

\begin{algorithm}[htbp]
\caption{CovertComBench Benchmark Construction Pipeline.}
\label{alg:covertcombench_pipeline}
\begin{algorithmic}[1]
\Require Raw source papers $\mathcal{P}_{\text{raw}}$
\Ensure Frozen benchmark $B_{\text{frozen}}$ with baseline scores and metadata

\Statex \textbf{Stage S1: Source Vetting \& Decontamination}
\State Filter papers based on peer review and relevance
\State Perform contamination check (TF-IDF + Sentence-BERT)
\State Collect clean papers $\rightarrow S_{\text{clean}}$ \Comment{QG1}

\Statex \textbf{Stage S2: Context Extraction \& Question Formulation}
\For{each paper $p \in S_{\text{clean}}$}
    \State Extract salient problem points (human-in-the-loop)
    \State Generate context snippets and classify into MCQ, ODQ, CGQ
    \State Normalize variables and validate units
    \State Add type-specific enhancements (distractors, rubrics, functional checks)
\EndFor
\State $D_{\text{questions}} \gets$ all generated draft questions \Comment{QG2}

\Statex \textbf{Stage S3: Instance Refinement \& Expert Validation}
\For{each question $q \in D_{\text{questions}}$}
    \State Standardize notation and enrich metadata
    \State Expert review: correctness, solvability, clarity
    \State Compute source similarity and fix hash
\EndFor
\State $V_{\text{questions}} \gets$ all validated questions \Comment{QG3}

\Statex \textbf{Stage S4: Finalization, Trial Run, \& Baselining}
\State Freeze dataset, compute dataset hash
\State Conduct trial runs with baseline models
\State Record performance scores and categorize errors \Comment{QG4}
\State \Return $B_{\text{frozen}}$
\end{algorithmic}
\end{algorithm}

To support fine-grained performance evaluation and reproducibility, the entire benchmark is stratified by difficulty and scenario. 
Each question is annotated with detailed metadata, including its scenario cluster, difficulty level, expected solution tag, and permissible equivalent derivation strategies, for subsequent analysis and reproduction.

\subsection{Question Structure and Design}
The core design principle of CovertComBench is to isolate the unique challenges of CC, distinguishing them from broader issues in general wireless communications. Its tasks, curated from a comprehensive survey of CC scenarios, are meticulously designed to probe a deep understanding of the field's central challenge: the fundamental trade-off between communication performance and covertness, and the strategic decisions this balance necessitates. In system performance analysis, LLMs are challenged not only to maximize the receiver's Signal-to-Noise Ratio (SNR) but to do so under the stringent constraint that a monitor's detection performance remains below a critical threshold.
We prioritize reasoning density over dataset size. While other benchmarks may contain thousands of shallow questions, they often lack the depth required to evaluate secure communication systems.
To facilitate a multi-layered evaluation, we designed three distinct question formats, each targeting a different dimension of a model's capabilities:

\begin{itemize}
    \item \textbf{MCQs} evaluate LLM's comprehension of core CC concepts and its ability to make judicious trade-off decisions under complex constraints.

    \item \textbf{ODQs} probe LLM's core reasoning capabilities, specifically its capacity for symbolic mathematical derivation and logical deduction when solving complex optimization problems.

    \item \textbf{CGQs} measure LLM's ability to translate theoretical models and mathematical formulas into executable code for quantitative analysis.
\end{itemize}
\begin{figure}
    \centering
    \includegraphics[width=0.8\linewidth]{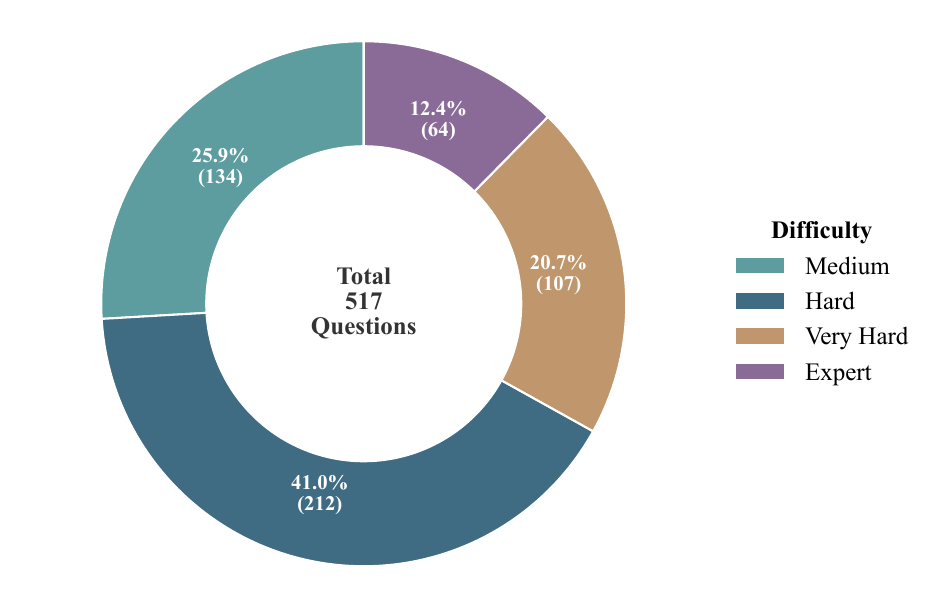}
    \caption{CovertComBench difficulty distribution.}
    \label{fig:difficulty_distribution}
\end{figure}
\begin{table}[htbp]
\centering
\caption{Detailed Statistics of Task Types and Difficulties in the Benchmark.}
\label{tab:task_distribution}
\footnotesize
\begin{tabularx}{\columnwidth}{@{}lXcc@{}}
\toprule
\textbf{Category} & \textbf{Task Type} & \textbf{Count} & \textbf{Percentage} \\
\midrule
\multirow{6}{*}{\textbf{MCQ}} 
 & Theoretical Analysis & 100 & 42.2\% \\
 & Optimization Path \& Result Selection & 63 & 26.6\% \\
 & Technical Design & 44 & 18.6\% \\
 & Scene Adaptation & 30 & 12.7\% \\
 \cmidrule(l){2-4}
 & \textit{Total} & \textit{237} & \textit{100.0\%} \\
\midrule
\multirow{6}{*}{\textbf{ODQ}} 
 & Closed-form Expression Derivation & 58 & 39.2\% \\
 & Optimization Problem Solving & 45 & 30.4\% \\
 & Optimization Algorithm Reasoning & 41 & 27.7\% \\
 & Scenario Adaptation Inference & 4 & 2.7\% \\
 \cmidrule(l){2-4}
 & \textit{Total} & \textit{148} & \textit{100.0\%} \\
\midrule
\multirow{6}{*}{\textbf{CGQ}} 

 & Complex Symbolic Computing& 57 & 43.2\% \\

 & Secrecy Performance Quantification & 43 & 32.6\% \\

 & Optimization Solving Implementation & 32 & 24.2\% \\
 \cmidrule(l){2-4}
 & \textit{Total} & \textit{132} & \textit{100.0\%} \\
\bottomrule
\end{tabularx}
\end{table}
By integrating these three question types, CovertComBench establishes a comprehensive and multifaceted benchmark. It moves beyond simple fact retrieval to systematically assess an LLM's ability to understand, apply, and reason about the complex principles of CC. Consequently, it provides a nuanced and structured assessment of an LLM's potential as a scientific assistant in this domain.

\subsection{Dataset Statistics and Splits}
To facilitate a comprehensive evaluation of CovertComBench, we have conducted detailed statistical analysis and annotation of the dataset. We have categorized the problem difficulty into four distinct levels: Medium, Hard, Very Hard, and Expert. We have designed four distinct task categories tailored to different types of problems. The distribution of these tasks among the various problem types is depicted in \Cref{tab:task_distribution}. The number of questions designated as Expert level is intentionally limited. \Cref{fig:difficulty_distribution} illustrates the distribution of questions across these difficulty levels within the dataset.

\section{Experiments}
This section presents a comprehensive evaluation of CovertComBench. We begin by detailing the experimental design, including the set of evaluated LLMs and the prompt engineering. To enhance the discrimination of different LLMs, we introduce tailored evaluation metrics. Supported by extensive experiments, our analysis reveals the performance characteristics of the evaluated models across diverse task categories and highlights the fundamental bottlenecks and key challenges faced by current LLMs in CC tasks.
\begin{table}[t]
\centering
\renewcommand{\arraystretch}{1.25}
\caption{LLMs list in this CovertComBench. The table details the evaluated LLMs, including their deployment method (API or Local) and parameter size (in billions).} 
\begin{tabular}{@{}lcccccc@{}}
\toprule
\textbf{Model} &
\textbf{Deployment Method} &
\textbf{Size}
\\ 
\midrule 

DeepSeek-V3.2-reasoner  & API & 671B \\
DeepSeek-V3.2-chat & API & 671B \\
OpenAI-o3  & API & Unknown \\
Gemini-3.0-Pro   & API & Unknown \\
Gemini-3.0-Flash  & API & Unknown \\
DeepSeek-R1 & Local & 70B \\
Llama3.3  & Local & 70B\\
Wizard-math & Local & 70B \& 7B\\

Qwen3 & Local & 32B\\
Gemma3 & Local & 27B\\
Magistral & Local & 24B\\
GPT-oss & Local & 20B\\
Llama3 & Local & 7B\\
Mistral & Local & 7B\\

\bottomrule
\end{tabular}

\label{tab:Evaluated Models} 
\end{table}
\subsection{Experiment Setup}
We conducted a comprehensive evaluation of a diverse suite of LLMs on the CovertComBench benchmark. The selected models, presented in \Cref{tab:Evaluated Models}, encompass a wide range of architectures and parameter scales, accessed via both local deployment and API calls. To manage the local models and ensure a consistent execution environment, we utilized the Ollama framework.

To ensure fair and reproducible evaluation, all tasks were formatted using a unified few-shot prompt template and standardized inference hyperparameters (temperature = 0.2, top-p = 0.95, max new tokens = -1, random seed = 42, repetition penalty = 1.1). For locally deployed models, these parameters were explicitly configured to maintain consistency, whereas partial API-based models, which do not allow hyperparameter definition, were evaluated under their respective default inference settings. This setup ensures that performance differences reflect the models’ intrinsic abilities rather than variations in the experimental configuration.

\begin{table*}[htbp]
\centering
\small
\renewcommand{\arraystretch}{0.9} 
\caption{Performance metrics of different LLMs across question types. MCQs and CGQs are evaluated using automated scripts, whereas ODQ scores are derived from manual evaluation by human experts (scored out of 1480, 10 points per question). The MAE (Mean Absolute Error) column represents the average discrepancy between the scores assigned by the LLMs and the human expert. Underlined values indicate outliers due to data contamination.}
\begin{tabular}{l rr rrr rr}
\toprule

& \multicolumn{2}{c}{\textbf{MCQs}} & \multicolumn{3}{c}{\textbf{ODQs}} & \multicolumn{2}{c}{\textbf{CGQs}} \\
\cmidrule(lr){2-3} \cmidrule(lr){4-6} \cmidrule(lr){7-8}

\textbf{Model} & \textbf{Accuracy (\%)} & \textbf{F1 Score} & \textbf{Accuracy (\%)} & \textbf{Score} & \textbf{MAE} & \textbf{Accuracy (\%)} & \textbf{Score } \\
\midrule
Gemini-3.0-Pro(API)    & 80.17  & \textbf{0.92} & 53.37  &1066 & \textbf{3.22}& \textbf{83.33}  & \textbf{1043} \\
Gemini-3.0-Flash(API)  & 68.35 & 0.81 & 39.86 & 857 &3.60& 78.03 & 964\\
DeepSeek-V3.2-reasoner(API)        & 72.57 & 0.88 & 48.65 &986 & 3.87 & 79.55& 1035\\
DeepSeek-V3.2-chat(API)        & 64.14& 0.82 & 36.49 & 822 & 4.10 & 68.94 & 853\\
OpenAI-o3(API)         & \textbf{81.86}& 0.91 & \textbf{55.41}& \textbf{1073} & 3.35 & 82.58 & 1024\\
\midrule
DeepSeek-R1:70B        & \textbf{70.46} & \textbf{0.84} & \textbf{43.91} & \textbf{890} & 4.13& \textbf{71.21} & \textbf{835}\\
Llama3.3:70B            &63.29 & 0.81 & 33.33 & 770 & \textbf{2.65} & 56.82 & 690\\
Wizard-math:70B        &16.46 & 0.46& 18.18 & 498 & 4.55 & 15.15 & 185\\
\midrule
Qwen3:32B               &\textbf{64.56} & \textbf{0.80} & \textbf{37.84} & \textbf{855} & \textbf{2.24} & \textbf{66.67} &\textbf{856}\\
Gemma3:27B             &53.59 & 0.75 & 28.38 & 765 & 3.24 & 62.12 &736\\
Magistral:24B          &42.19 & 0.70 & 26.35 & 522 & 3.84 & 59.85 &718\\
GPT-oss:20B            &\textbf{\underline{72.80}} & \textbf{\underline{0.89}} & \textbf{\underline{39.86}} &  \textbf{\underline{783}} & 4.35 & \textbf{\underline{70.45}}& \textbf{\underline{897}}\\
\midrule
Llama3:8B              &\textbf{44.30}& \textbf{0.79}& \textbf{23.65} & \textbf{525} & 4.12 & \textbf{53.79} & \textbf{668}\\
Mistral:7B             &36.29&0.64 &22.30 &503 & \textbf{3.64} & 43.94 & 526\\
Wizard-math:7B         &27.42 &0.49 & 18.18 & 435 & 4.75 & 17.42 &200\\
\bottomrule
\end{tabular}
\label{tab:performance_metrics_accuracy}
\end{table*}

\subsection{Evaluation Metrics}

\textbf{MCQs:} Our evaluation employs two distinct metrics to provide a comprehensive assessment of model performance: exact match accuracy and sample-averaged F1 score.

\textbf{ODQs :} ODQs often involve optimization and tradeoff analysis; the evaluation focuses on the soundness of the reasoning process rather than solely on the final answer. To formalize this, we adopt a process-centric scoring mechanism.

Each ODQ is broken down into $K$ predefined reasoning checkpoints. For each checkpoint $k$, we define an indicator function $l_k \in [0, 1]$ that represents the correctness of that specific reasoning step, and assign it a corresponding weight $w_k > 0$ such that $\sum_{k=1}^{K} w_k = 1$. The process score, $S_{\mathrm{proc}}$, is then calculated as the weighted sum of the scores of all checkpoints:
\begin{equation}
S_{\mathrm{proc}} = \sum_{k=1}^{K} w_k \cdot l_k
\end{equation}
Additionally, to account for the correctness of the final closed-form expression or numerical answer, we introduce an indicator $I_{\mathrm{ans}}$ (1 if correct, 0 otherwise). The final score for an ODQ, $S_{\mathrm{ODQ}}$, is a linear interpolation between the process score and the final answer correctness, controlled by a hyperparameter $\lambda \in [0, 1]$:
\begin{equation}
S_{\mathrm{ODQ}} = \lambda S_{\mathrm{proc}} + (1-\lambda) I_{\mathrm{ans}}
\end{equation}
The value of $\lambda$ is recorded in each question's metadata, allowing for flexibility in balancing process versus outcome. To ensure fairness when evaluating logically equivalent derivations or numerical results, we employ techniques such as normalized variable mapping, structural matching, and relative tolerance (rtol) for numerical comparisons.

\textbf{CGQs:} To evaluate the engineering quality of the code generated by LLMs, we employ a multi-turn iterative testing approach. Initially, the task description is provided to the LLM. Subsequently, the generated code is executed within a test script. We define a ``failure'' as any instance where the code encounters runtime exceptions or produces outputs that do not match the expected results. In the event of a failure, the specific error message or assertion feedback is returned to the LLM as a rectification prompt. This cycle repeats until the code passes all tests or the maximum number of attempts ($N_{\max}=3$) is exceeded.

The final score, $S_{\mathrm{CGQ}}$, is determined by a decay function based on the number of attempts, denoted as $k$, required to obtain a correct solution. We prioritize ``one-shot'' success while awarding partial credit for successful debugging within a limited window. The scoring function is defined as:

\begin{equation}
S_{\mathrm{CGQ}} = \begin{cases} 
10, & \text{if } k = 1 \ (\text{One-shot success}) \\
7, & \text{if } k = 2 \ (\text{Success after 1st revision}) \\
4, & \text{if } k = 3 \ (\text{Success after 2nd revision}) \\
0, & \text{if } k > 3 \ (\text{Failure})
\end{cases}
\label{eq:cgq_score}
\end{equation}

\subsection{Experiment Result}
\Cref{tab:performance_metrics_accuracy} summarizes the performance of various LLMs across the three tasks. The results illustrate significant trends regarding both solution quality and evaluation reliability. 

First, there is a clear discrepancy between task types: most models, particularly advanced API-based ones, demonstrate strong proficiency in MCQs and CGQs. However, performance drops in ODQs, indicating that current LLMs struggle with the rigorous mathematical reasoning required for CC. \textbf{This difficulty extends to evaluation, as reflected in the MAE metrics.} Unlike human experts who assign precise partial credit based on reasoning checkpoints, LLM-based judges tend to exhibit polarized scoring behaviors—often significantly overscoring or underscoring solutions. This lack of granularity leads to a notable divergence from human ground truth in measuring mathematical derivation.

Second, the impact of training data quality is evident. For instance, GPT-oss achieves exceptionally high scores relative to its parameter size, a result attributed to the inclusion of domain-specific academic literature in its training corpus.

\subsection{Main Error Analysis}\textbf{Semantic Misalignment and Distractor Vulnerability:} LLMs are susceptible to "distractors" in MCQs, often validating propositions based on partially correct segments while overlooking errors. Furthermore, we observed an embedding space misalignment: models conflate ``Covert Communication'' with multimedia ``Steganography'' (e.g., image hiding) rather than wireless physical layer attributes (e.g., signals, power). This results in fundamental conceptual bias.

\textbf{Failures in Non-Algorithmic Symbolic Calculation:} While LLMs handle simple differentiation via pattern matching, they struggle with integration and expectation calculations crucial for CC. Unlike differentiation, integration demands strategic substitution and deep functional understanding—capabilities that remain elusive for models relying on statistical pattern matching.

\textbf{Optimization Bias and Security Constraint Violation:} Models exhibit a strong bias towards maximizing utility (e.g., transmission rate) while neglecting hard constraints. We observed correct final answers derived through processes that explicitly ignored covertness limits. In security-critical contexts, such negligence renders the solution unsafe despite maximized utility.

\textbf{Persistent Hallucinations in Code Generation:} Library hallucinations remain persistent; LLMs frequently invoke non-existent functions. Notably, models often regenerate identical erroneous code even after receiving explicit error feedback (e.g., ``AttributeError''), indicating an inability to self-correct hallucinations via simple prompting.

\textbf{Pathways to Improvement:} We propose three strategies: 
1) \textbf{Tool-Augmented Fine-Tuning}: Shift calculation reliance from internal weights to external tools (e.g., Function Calling). Training data must include tool documentation to minimize hallucinations. 
2) \textbf{Negative Sample Training}: Since standard fine-tuning on correct data is insufficient, datasets should incorporate plausible but incorrect derivations to enhance discriminative capabilities. 
3) \textbf{Closed-Loop Feedback Agents}: Implement iterative agents that interpret execution errors and cross-reference libraries to enable dynamic debugging rather than repetitive regeneration.
\section{Conclusion}
In this paper, we introduced CovertComBench, the first systematic benchmark designed to evaluate LLMs in CC systems. Our comprehensive evaluation reveals a critical performance dichotomy: LLMs demonstrate strong capabilities in conceptual understanding and code generation but show significant limitations in tasks demanding rigorous, multi-step mathematical reasoning. This leads to our central conclusion that current LLMs function as highly efficient auxiliary tools rather than as autonomous problem-solvers in this specialized field. To advance their utility, future research must focus on enhancing their problem decomposition, step-by-step reasoning, and domain-specific fine-tuning. By addressing these fundamental challenges, LLMs can evolve from useful assistants into more capable partners in scientific research. Meanwhile, we will continuously update and extend CovertComBench to ensure the quantity and quality of the dataset.
\bibliography{references}
\bibliographystyle{IEEEtran}
\newpage
\end{document}